\documentclass[useAMS,usenatbib]{mn2e}

\usepackage[psamsfonts]{amssymb}
\usepackage[dvips]{graphicx}
\usepackage{amsmath,alltt}
\usepackage{multirow}
\usepackage{rotating}
\usepackage{lscape}

\title[The state of globular clusters at birth]{The state of globular clusters at birth: emergence 
from the gas-embedded phase}
\author[Leigh et al.]{Nathan Leigh$^{1}$, 
Mirek Giersz$^{2}$, Jeremy J. Webb$^{3}$, Arkadiusz Hypki$^{2}$, Guido De Marchi$^{1}$, 
\newauthor 
Pavel Kroupa$^{4}$, Alison Sills$^{3}$
\thanks{E-mail: nleigh@rssd.esa.int (NL); mig@camk.edu.pl (MG); webbjj@univmail.cis.mcmaster.ca (JW); 
ahypki@camk.edu.pl (AH); gdemarchi@rssd.esa.int (GD); pavel@astro.uni-bonn.de (PK); asills@mcmaster.ca (AS)}\\
$^{1}$European Space Agency, Space Science Department, Keplerlaan 1,
2200 AG Noordwijk, The Netherlands \\
$^{2}$Nicolaus Copernicus Astronomical Centre, Polish Academy of Sciences, ul. Bartycka 18, 00-716 Warsaw, Poland \\
$^{3}$McMaster University, Department of Physics and Astronomy, 1280 Main St. W., 
Hamilton, Ontario, Canada, L8S 4M1 \\
$^4$Helmholtz-Institut f\"ur Strahlen- und Kernphysik, Nussallee 14-16, D-53115 Bonn, Germany}
\begin{document}

\pagerange{\pageref{firstpage}--\pageref{lastpage}} \pubyear{2011}

\maketitle

\label{firstpage}

\begin{abstract}


In this paper, we discuss the origin of the observed correlation between cluster 
concentration c and present-day mass function (PDMF) slope $\alpha$ reported by De Marchi, 
Paresce \& Pulone.  This relation can either be reproduced from 
universal initial conditions combined with some dynamical mechanism(s) that alter(s) the 
cluster structure and mass function over time, or it must arise early on in the cluster 
lifetime, such as during the gas-embedded 
phase of cluster formation.  Using a combination of Monte Carlo and $N$-body models for 
globular cluster evolution performed with the MOCCA and NBODY6 codes, respectively, we 
explore a number of dynamical mechanisms that could affect the observed relation.  

For the range of initial conditions considered here, our results are consistent 
with an universal initial binary fraction $\approx$ 10\% (which does not, however, 
preclude 100\%) and an universal initial stellar mass function resembling the standard Kroupa 
distribution.  \textit{Most of the dispersion observed in the c-$\alpha$ relation 
can be attributed to two-body relaxation and Galactic tides.  However, dynamical processes 
alone could not have reproduced the dispersion in concentration, and we require at least 
some correlation between the initial concentration and the total cluster mass.}  
We argue that the origin of this trend 
could be connected to the gas-embedded phase of cluster evolution.  

\end{abstract}


\section{Introduction} \label{intro}

\citet{demarchi07} showed that Milky Way (MW) globular 
clusters (GCs) exhibit a correlation between the logarithmic ratio of 
their tidal $r_t$ and core $r_c$ radii, called the concentration 
parameter $c =$ log($r_t/r_c$), and the slope of the low-mass stellar 
global mass function (MF) $\alpha$.  That is, high concentration clusters 
tend to have steep MFs, while low concentration clusters tend to have 
flat MFs.  The authors posited that this goes against the naive 
expectation that it is solely two-body relaxation that drives the 
evolution of both the concentration and the MF slope.  In particular, 
two-body relaxation causes the preferential evaporation of low-mass 
stars across the tidal boundary, while at the same time driving clusters 
toward a state of higher central density \citep[e.g.][]{spitzer87,heggie03}.  
It follows that more concentrated clusters should be more severely 
depleted of preferentially low-mass stars and have a shallower MF than 
low-concentration clusters.  This is precisely the opposite of what was 
found by \citet{demarchi07} for a sample of 20 MW GCs.  Clearly, the 
observed correlation between concentration and MF slope is puzzling.

Several authors have suggested mechanisms to explain this curious trend.  
\citet{demarchi07} offered that GCs severely depleted of their low-mass 
stars underwent core collapse some time in the past, and have since 
recovered a normal radial density profile.  However, this cannot 
explain the high concentrations observed in very massive clusters.  
Alternatively, \citet{marks08} suggested that the 
observed correlation could be due to residual gas explusion from 
initially mass segregated clusters, combined with the effects of 
unresolved binaries.  
%
The authors argued that extreme gas expulsion could produce clusters with 
low central densities, and hence low concentrations, and flat 
PDMFs at the low-mass end.  
Studying the interplay between the stellar initial mass function (IMF), 
concentration, and gas retention plays a crucial role in understanding 
the origins of star clusters in the Milky Way.  For example, this 
approach leads to constraints on the sequence of events during the 
early evolution of the proto-Galaxy \citep{marks10}, on 
the variation of the IMF in star bursts \citep{marks12} and on the 
initial mass-radius relation of star clusters \citep{murray09,marks12b}, 
and on the observed distribution of GC metallicities \citep{marks10}.

\citet{leigh12} showed that the cluster-to-cluster variations observed 
in the PDMFs of MW GCs are consistent with what is expected 
if GCs were born with a universal initial mass function (IMF), and 
two-body relaxation is the dominant physical process driving the 
evolution of the MF.  In other words, the power-law index of the 
central MF $\alpha$ increases (i.e. the 
MF flattens) smoothly with decreasing cluster mass.  This is what is 
expected from two-body relaxation, since it segregates low-mass stars to 
the outskirts where they escape from the cluster across the tidal 
boundary, and it operates with a rate that increases 
with decreasing cluster mass.  This suggests that whatever mechanism 
is responsible for the observed correlation between $c$ and $\alpha$, 
it should have primarily affected the cluster concentration relative to 
what is expected from two-body relaxation alone, as opposed to the 
slope of the MF $\alpha$.  Said another way, if two-body relaxation is the 
only mechanism driving the internal evolution of clusters, then the expected 
dependence of $\alpha$ on the total cluster mass is roughly consistent with what 
is observed.\footnote{This assumes that all clusters were born with similar 
initial mass functions, and is based on the central stellar MF, which is 
relatively insensitive to the effects of Galactic tides.}  This is not the 
case for the cluster concentration, however.

Other observational correlations have been reported for MW GCs, some of which are
arguably also consistent with the 
general picture that GCs were born with approximately universal initial
conditions and evolved via dynamics to their present-day forms.  For example,
\citet{milone12} recently performed a detailed study of the properties of
main-sequence binaries in a sample of 59 GCs.  The authors confirmed a
previously reported \citep{sollima07} anti-correlation between the binary
fraction and the total cluster mass.  \citet{sollima08} showed via analytic
methods that such an anti-correlation can arise naturally assuming an
universal initial binary fraction that is independent of cluster mass.
This can be explained by the disruption of binaries in the cluster core, 
combined with the evaporation of single stars from the cluster outskirts 
\citep[e.g.][]{fregeau09}.  The efficiency of the former should increase
with increasing cluster density \citep{marks11} and hence mass, whereas the 
efficiency of the latter is driven 
by two-body relaxation and should increase with decreasing cluster mass.  
This contributes to high binary fractions in low-mass clusters, and low 
binary fractions in high-mass clusters.  Notwithstanding, 
\citet{sollima08} cautioned that, based on existing observations
of binary fractions in Galactic GCs, the data is also consistent with
significant variations among the initial binary properties.  This would, 
however, contradict the well-motivated universality hypothesis \citep{kroupa11a}.

In this paper, we argue that the origin of the observed distribution of concentration 
parameters $c$ in GCs must be connected to 
the gas-embedded phase of cluster formation.  The alternative 
is that globular clusters emerged from the embedded phase with universal initial 
conditions, and evolved via dynamics to their presently observed 
MFs and structural parameters.  We perform a suite of numerical 
simulations for comparison to the observed c-$\alpha$ relation, varying the 
initial conditions in each model in order to identify those that yield the 
best agreement with the observations at the present-day cluster age.  
We also compare the simulated distribution of binary fractions as a function of
the total cluster luminosity to the observed relation of \citet{milone12} in 
an effort to constrain to first-order the universality of the initial 
binary fraction in GCs.

We begin by considering in Section~\ref{dyn-sinks} different initial conditions 
that could affect the evolution of the MF slope and/or the concentration parameter.  
In Section~\ref{models}, we describe the specific initial conditions we consider, 
as well as the Monte Carlo and $N$-body models used to simulate the cluster 
evolution.  We present our results in Section~\ref{results} and, based on 
these results, we argue in Section~\ref{discussion} that 
dynamics alone could not have reproduced the observed $c-\alpha$ relation.  
This implies that the observed relation must have originated very early 
on in the cluster lifetime, when gas was still present in 
significant quantities.  Hence, we 
discuss the various mechanisms that could have contributed to the observed 
distribution of concentration parameters during the gas-embedded phase, and 
constrain the necessary conditions.  We conclude in Section~\ref{summary}.


\section{Dynamical mechanisms affecting the $\MakeLowercase{c}$-$\alpha$ relation} \label{dyn-sinks}

In this section, we consider several different dynamical mechanisms 
that could affect the evolution of the MF slope and/or the cluster 
concentration.  We further describe the initial conditions for which 
each of these mechanisms should contribute to the observed c-$\alpha$ 
relation.

\subsection{Binary stars} \label{binaries}

\subsubsection{Soft binaries} \label{soft-bin}


Soft binaries are characterized by their orbital energy, which must 
have an absolute value that is less than the average single star kinetic 
energy.  This inequality gives \citep{heggie75}:
\begin{equation}
\label{eqn:hard-soft}
a_{\rm soft} > \frac{G\bar{m}}{\sigma^2},
\end{equation}  
where a$_{\rm soft}$ denotes the semi-major axis of a soft binary, $\bar{m}$ is 
the average stellar mass, and $\sigma$ is the velocity 
dispersion.  If a soft binary experiences a direct encounter 
with a single star, the total energy of such an encounter is positive, and the 
binary will likely be disrupted.  Thus, on average, the disruption of a soft binary 
by a single star serves to reduce the interloper's speed, and hence kinetic 
energy.

If enough soft binaries are disrupted, this could affect the 
distribution of stellar velocities in a cluster.  The cooling of a cluster through 
binary star disruption was first demonstrated by \citet{kroupa99}.  More recently, 
\citet{fregeau09} showed that the initial energy in soft binaries can be 
up to 10\% of the total mechanical energy of a cluster.  The authors
argued that this is a sufficiently significant energy sink to drive a typical 
MW GC to core collapse.  The disruption of soft binaries also causes 
a decrease in the binary fraction, not only due to the destruction of binaries but 
also due to the increase in single stars.  This sudden increase in the number of 
single stars further contributes to increasing the relaxation time by increasing the 
total number of objects and decreasing the average object mass.  Having said 
that, most soft binaries are disrupted very early on in the cluster lifetime, when 
massive stars are still present.  The stellar evolution-driven mass loss 
from massive stars contributes to heating the core, causing the core radius to 
expand.  The question is: does this tend to outweigh the energy sink provided by 
soft binaries, so that their disruption only serves to slow the expansion of the core?

The semi-major axis corresponding 
to the hard-soft boundary decreases with increasing cluster mass, since the velocity 
dispersion increases with increasing cluster mass (see Equation~\ref{eqn:hard-soft}).  
Thus, assuming an universal initial binary fraction that is independent of the cluster 
mass, more massive clusters have more soft binaries initially 
\citep{kroupa95,marks11}.  It follows 
that the efficiency of soft binary disruption as an energy sink should increase 
with increasing cluster mass.  In other words, for a universal initial 
binary fraction and orbital parameter distributions, the disruption of 
soft binaries should contribute to a correlation between the 
total cluster mass and the concentration parameter, as observed.

\subsubsection{Hard binaries} \label{hard-bin}

Hard binaries, for which the absolute value of the orbital energy exceeds the 
average kinetic energy of a single star, can also influence the central 
concentration via ``binary burning'' \citep[e.g.][]{fregeau09}.  In this case, 
the central density is sufficiently high that even very close binaries, for which 
the collisional cross-section is small, frequently undergo dynamical interactions 
with single stars.  Here, the binary imparts additional kinetic energy to the 
escaping single star, becoming even harder in the process \citep{heggie75}.  Thus, 
hard binaries can act as heat sources in clusters with sufficiently high central 
densities, slowing and even reversing the tendency toward core collapse 
\citep[e.g.][]{heggie03}.  

For a given binary fraction, low-mass clusters should contain the largest 
number of hard binaries, since the hard-soft boundaries in these clusters correspond 
to large orbital separations.  Hence, core expansion driven by hard binary burning 
should also contribute to a correlation between the cluster mass and concentration.  
\citet{fregeau09} argued that 
most Milky Way GCs have not yet reached sufficiently high central densities to enter 
the binary burning phase of evolution, although the authors assumed 
that the concentration was lower in the past and that it has been increasing steadily 
over time.  Clusters with very high initial concentrations, on the other hand, are more 
likely to undergo binary 
burning early on, and this can contribute to decreasing the concentration parameter 
\citep{heggie08,heggie09}.  

\subsection{Cluster expansion in a tidal field} \label{exp-tidal}

Star clusters expand self-similarly in a tidal field \citep{gieles11}.  This 
contributes to decreasing the concentration parameter or, more accurately, 
reducing the rate at which the concentration increases due to two-body 
relaxation.  This effect occurs in clusters that are 
initially tidally under-filling, since they have room to expand and 
experience a long delay before losing mass across the tidal boundary.  Once a 
cluster fills its tidal radius, it experiences a steady increase in its 
concentration as the core radius shrinks due to two-body relaxation.

Assuming a universal distribution of initial cluster sizes, it is the most massive 
clusters that should expand the most in an under-filled tidal field, since 
they have the largest tidal radii.  A useful, albeit simplistic \citep{webb12,webb13}, 
approximation for the tidal radius is \citep{vonhoerner57}:
\begin{equation}
\label{eqn:r_t}
r_{\rm t} = R_{\rm GC}\Big( \frac{M_{\rm clus}}{M_{\rm g}} \Big)^{1/3},
\end{equation}
where $R_{\rm GC}$ is the Galactocentric distance of the cluster assuming a 
circular orbit, $M_{\rm clus}$ is the cluster mass, and $M_{\rm g}$ is the mass of 
the Galaxy, which here we assume to be a point mass.  

Two-body relaxation acts to move lower mass stars outward to the cluster outskirts, 
where they preferentially escape across the tidal boundary.  A cluster that 
experiences a stronger tidal field will undergo a more rapid stripping of 
its low-mass stars, due to the deeper potential in which the cluster sits and also 
the fact that a stronger tidal field translates into a smaller 
tidal radius.  As the cluster loses mass, the process is accelerated, since 
the tidal radius decreases and the rate of two-body relaxation increases.  For 
clusters with small Galactocentric distances, this can 
even result in an inversion of the mass function, so that its slope changes 
sign \citep{vesperini97,baumgardt03}.

Tidal heating acts only on clusters with eccentric orbits, since a non-static 
tidal field implies that the depth of the gravitational potential is periodic, and 
additional energy is deposited within the stellar population with each 
perigalactic pass.  This accelerates the rate 
of escape of preferentially low-mass stars across the tidal boundary.  This 
is because low-mass stars in the outskirts have the lowest binding energies, 
so the additional energy accelerates these stars to speeds that exceed the 
escape velocity, at which point they have a positive total 
energy and become unbound.  Overall, the net effect of tidal heating is to 
accelerate stars, causing an expansion of the cluster and its more rapid 
dissolution.  Importantly, the mean mass loss per unit time due to external 
perturbations from the Galaxy are independent of cluster mass 
\citep{gieles06,gieles07}.  Thus, 
tidal heating should contribute to the observed dependence 
of mass function slope on concentration, since, for a given orbit, it should 
produce more extended clusters with shallower MF slopes for smaller initial 
cluster masses, in agreement with the observations.


\subsection{Initial stellar and remnant mass functions} \label{remnants}

The initial mass function can affect the evolution of the cluster structure in a 
number of ways.  For example, for a given cluster mass, the initial mass function  
determines both the total number of stars and the average stellar mass, and hence 
the rate of two-body relaxation.  Mass loss due to stellar evolution tends
to cause clusters to expand \citep[e.g.][]{chernoff90}.  Thus, clusters with 
a top-heavy initial mass function should expand more due to stellar evolution than 
clusters with a bottom-heavy initial mass function, since the former includes a 
larger fraction of 
massive stars.  A top-heavy IMF should also generate more 
stellar remnants early on in the cluster lifetime, since the stellar lifetime 
decreases with increasing stellar mass.

Stellar remnants, including white dwarfs, neutron stars and black holes, 
represent an unseen component of globular clusters.  Assuming a standard 
initial mass function \citep{kroupa02}, they could constitute 
a substantial fraction of the total cluster mass at an age of $\approx 12$ Gyr 
\citep[e.g.][]{leigh13}.  More 
importantly, remnants make up a much larger fraction of the \textit{core} 
mass, since they are primarily confined to the central cluster region.  This 
is because, at the time of their formation, stellar remnants descend from the 
most massive stars in the cluster, which have typically migrated into the core 
via mass segregation by the time of their death, if they did not form their in 
the first place.  Even in very massive clusters for which the rate of two-body 
relaxation and hence mass segregation is slower, remnants that do not form in the 
core will still quickly migrate there via two-body relaxation since, after a few 
hundred Myr, they are the most massive objects in the cluster.

Given that they represent a significant, but unseen, mass component within 
the core, stellar remnants can have an important bearing on the concentration 
parameter.  In particular, remnants should act as an additional heat source 
within the core, and we expect this to contribute to an increase in the core 
radius, and hence a decrease in the concentration parameter 
\citep[e.g.][]{mackey07,mackey08,sippel13}.  

Stellar remnants could contribute (weakly) to the observed c-$\alpha$ relation.  
This is because, 
for a larger cluster mass, more massive stars are more likely to be selected from 
random sampling, and there is evidence that the maximum IMF mass increases with 
increasing cluster mass \citep{kroupa13}.  More massive stars implies more massive 
remnants, which are more effective at heating the core, and hence lowering the 
concentration.  It follows that, for a given IMF, stellar remnants could be 
more effective heat sources in more massive clusters, contributing to an 
anti-correlation between the cluster mass and concentration.  Thus, although 
the data are consistent with a universal IMF in Milky Way globular clusters 
\citep[e.g.][]{demarchi10,paust10,kroupa13,leigh12}, the stellar IMF could compete 
against the observed c-$\alpha$ relation indirectly via stellar remnants.  We do 
not expect this effect to typically be significant, however, given its stochastic 
nature, as reflected in the construction of the IMF via random sampling.

One interesting example that depends sensitively on the initial stellar and remnant 
mass functions involves the 
formation of an intermediate-mass black hole (IMBH).  An IMBH can form from the 
runaway collisions of massive stars very early on in the cluster lifetime 
\citep{portegieszwart04}.  The IMBH can then continue to grow via subsequent 
mergers, and even gas accretion mediated by binary star evolution.  If an IMBH forms, 
its presence should contribute to increasing the core radius, and hence 
decreasing the concentration parameter, by 
accelerating stars in its immediate vicinity \citep[e.g.][]{lutzgendorf11}.  If 
IMBHs in GCs follow a similar scaling law as super-massive black holes (SMBHs) in 
galactic nuclei \citep{lutzgendorf13,kruijssen13}, then we would expect more massive 
clusters to harbour more massive 
IMBHs.  In this case, IMBHs would contribute to a trend in which the concentration 
parameter decreases with increasing cluster mass, in clear disagreement with 
the observations.

\section{Models} \label{models}

In this section, we describe the Monte Carlo and $N$-body codes used 
to simulate the cluster evolution, and list the initial conditions 
we consider.  We use the Monte Carlo code MOCCA to simulate the majority 
of our model clusters, given its fast and robust coverage of the 
relevant parameter space.  The agreement between MOCCA and $N$-body models 
is excellent for the case of static Galactic tides \citep{giersz13}, however MOCCA cannot treat 
non-static tides.  Thus, we use the $N$-body code NBODY6 to quantify the 
impact of Galactic tides on the observed $c-\alpha$ relation, since it 
incorporates a realistic treatment of the Galactic potential.


\subsection{Monte Carlo models:  MOCCA} \label{mocca}

We use the MOCCA code to produce the majority of our simulated 
clusters.  It combines the Monte Carlo technique for cluster evolution 
with the \textit{Fewbody} code \citep{fregeau04} to perform numerical 
scattering experiments of small-number gravitational interactions, and 
relies on the Binary Stellar Evolution (BSE) code to track both stellar 
and binary evolution \citep{hurley00,hurley02}.

The MOCCA code offers 
several advantages, in particular the fast computation times required 
to run the simulations to completion.  It also allows us to simulate 
clusters in regions 
of parameter space that are intractable with $N$-body codes.  Specifically, 
since it relies on Monte Carlo methods, it can simulate realistic 
globular clusters composed of more than a million stars, and it can 
do so for 12+ Gyr of cluster evolution on time-scales of hours in 
real-time.\footnote{The MOCCA simulations are 
performed on a PSK2 cluster at the Nicolaus Copernicus Astronomical Centre in 
Poland.  Each simulation is run on one CPU.  The cluster is based on AMD (Advanced 
Micro Devices, Inc.) 
Opteron processors with 64-bit architecture (2-2.4 GHz).  For 12+ Gyr of 
GC evolution, simulations with 
5 $\times$ 10$^4$ stars typically take $\sim$ 1 hour to complete, 
those with 3 $\times$ 10$^5$ stars take 10-16 hours, 
and those with 1.8 $\times$ 10$^6$ stars take 120-160 hours.  The precise 
simulation run-time depends on the choice of initial conditions.}  For more 
detailed information about the MOCCA code, see \citet{hypki13} and 
\citet{giersz13}.

\subsubsection{Initial conditions} \label{initial}

We assume a King density profile with initial concentration W$_{\rm 0} = 6$.  
Our clusters are not mass segregated to begin with, and 
all models are initially tidally under-filling.  The degree of 
under-filling is set by the parameter f$_{\rm und} =$ r$_{\rm t}$/r$_{\rm h}$, 
where r$_{\rm t}$ and r$_{\rm h}$ are the tidal and half-mass radii, 
respectively.  To put this parameter into context, a King model \citep{king66} 
with W$_{\rm 0} = 6$ has a ratio between the tidal and half-mass radii of 
f$_{\rm und} = 6.79$.  We adopt a metallicity of $Z = 0.001$ for all models.

The initial conditions we consider are shown in Table~\ref{table:initial-MC}.  
We vary the initial mass function, binary fraction, kick velocities for NSs and 
BHs upon formation, the initial-final mass relation for BHs, the total number 
of stars, the ratio r$_{\rm t}$/r$_{\rm h}$, and the cluster age.  We run 
models having a total of 5 $\times$ 10$^4$, 10$^5$, 2 $\times$ 10$^5$, 
3 $\times$ 10$^5$ or 1.8 $\times$ 10$^6$ stars initially.  For the 
$N = 5 \times 10^4$ and $N = 3 \times 10^5$ cases, we also re-run identical 
models with the same initial conditions but different random number seeds to 
assess fluctuations in the final cluster state that are intrinsic to the 
Monte Carlo method adopted by MOCCA.  Snapshots are taken at 10, 11 and 12 
Gyr for all models (unless indicated otherwise).

We adopt two different IMFs, both in the form:
\begin{equation}
\label{eqn:imf}
\frac{dN}{dm} = m^{-\alpha}.
\end{equation}
The first IMF we refer to as a standard Kroupa IMF (labeled imf1 in 
Table~\ref{table:initial-MC}), as taken from \citet{kroupa93} in the 
mass range 0.08 - 100 M$_{\odot}$. The second IMF is a two-segment Kroupa IMF 
(imf2), as taken from \citet{kroupa08} 
with a single break mass at 0.5 M$_{\odot}$, and low- and high-mass slopes 
of +1.3 and +2.3, respectively.  In addition, we consider a modified IMF (imf3; 
instead of the two-segment Kroupa IMF) for 
the $N=100000$ and $N=200000$ cases, with a break mass at 0.85 M$_{\odot}$ and 
low- and high-mass slopes of +1.1 and +2.5, respectively.

For our standard model, the initial mass function adopted for the binaries is taken 
from Equation 1 of \citet{kroupa91} in the mass range 0.08 to 100 M$_{\odot}$, sampled 
with random pairing.  We assume different 
initial binary fractions of 10\%, 30\%, 70\% and 95\%, along with different 
maximum orbital separations of 100 AU, 200 AU and 400 AU.  
The binary semi-major axis 
distribution is uniform in the logarithmic scale from 2(R$_{\rm 1}$+R$_{\rm 2}$) 
to 100 AU (for the standard model).  The binary eccentricities follow a modified 
thermal distribution taken from Equation 1 of \citet{hurley05}.  We also 
perform simulations with the initial binary orbital parameter distributions provided 
in Equation 4.46 of \citet{kroupa13}, which are derived from empirical data (labeled 
Kroupa13 in Table~\ref{table:initial-MC}), in 
order to quantify the degree to which our assumptions for the initial 
binary orbital parameter distributions could affect our results.

Analytic formulae for stellar evolution are taken from \citet{hurley00}, and 
binary evolution is performed with the BSE code \citep{hurley02}.  
We adopt two different inital-final mass relations for BHs.  The first uses the 
initial-final mass relation from \citet{hurley00} assuming no mass fallback, and 
the second is the same but with mass fallback switched on \citep{belczynski02}.  
With mass fallback 
switched off, we adopt either a kick velocity of 265 km/s for both NSs and BHs 
(kick1), or we adopt 0 km/s for BHs and 265 km/s for NSs (kick2).  With mass 
fallback switched on, both the BH mass and kick velocity depend on the 
progenitor mass.

\begin{table*}
\caption{Initial conditions for all Monte Carlo (MOCCA) models.}
\begin{tabular}{|c|c|c|c|c|c|c|c|}
\hline
Total Number   &      Time     &  r$_{\rm t}$  &  f$_{\rm und}$  &  Binary    & a$_{\rm max}$ &   Model   &   Symbol   \\
of Stars       &    (in Gyr)   &   (in pc)     &                 &  Fraction  &    (in AU)    &           &            \\ 
\hline
1800000        &    10, 11, 12    &   125.33  &   60    &   10  & 100 &  standard (imf1 + kick1 + fallback)  & 5 pt. solid black triangle \\
               &                  &           &         &       &     &  no fallback   & 5 pt. solid blue triangle \\
               &                  &           &         &       &     &  imf2     & 5 pt. solid red triangle  \\
               &                  &           &   75    &       &     &  standard    & 5 pt. solid black square  \\
               &                  &           &         &       &     &  no fallback  & 5 pt. solid blue square  \\
               &                  &           &         &       &     &  imf2  & 5 pt. solid red square  \\
               &                  &           &         &       &     &  kick2  & 5 pt. solid green square \\
               &                  &           &         &       &     &  no fallback + imf2  & 5 pt. solid magenta square  \\
               &                  &           &         &       &     &  kick2 + imf2  & 5 pt. solid cyan square \\
               &                  &           &   75    &       &     &  Kroupa13  & 5 pt. blue cross  \\
300000         &    10, 11, 12    &    69.0   &   35    &   10  & 100 &  standard   & 4 pt. solid black triangle  \\
               &                  &           &         &       &     &  no fallback   & 4 pt. solid blue triangle  \\
               &                  &           &         &       &     &  imf2  & 4 pt. solid red triangle  \\
               &                  &           &         &       &     &  binary mass segregation  & 5 pt. black cross  \\
               &                  &           &         &   30  & 100 &  standard  & 1 pt. open black square  \\
               &                  &           &         &       & 200 &  standard  & 3 pt. open black square  \\
               &                  &           &         &       & 400 &  standard  & 5 pt. open black square  \\
               &                  &    38.0   &   35    &   70  & 100 &  standard  & 1 pt. open blue triangle  \\ 
               &                  &    69.0   &   35    &   70  & 100 &  standard  & 1 pt. open blue square  \\
               &                  &           &         &       & 400 &  standard  & 5 pt. open blue square  \\
               &                  &           &         &   95  & 100 &  standard  & 1 pt. open green square  \\
               &                  &           &         &       & 200 &  standard  & 3 pt. open green square  \\
               &                  &           &         &       & 400 &  standard  & 5 pt. open green square \\
               &                  &    69.0   &   50    &   10  & 100 &  standard  & 4 pt. solid black square  \\
               &                  &           &         &       &     &  no fallback  & 4 pt. solid blue square  \\
               &                  &           &         &       &     &  imf2  & 4 pt. solid red square  \\
               &                  &           &         &       &     &  kick2  & 4 pt. solid green square  \\
               &                  &           &         &       &     &  no fallback + imf2  & 4 pt. solid magenta square  \\
               &                  &           &         &       &     &  kick2 + imf2  & 4 pt. solid green square  \\
               &                  &           &   65    &       &     &  standard  & 4 pt. solid black pentagon  \\
               &                  &           &         &       &     &  no fallback  & 4 pt. solid blue pentagon  \\
               &                  &           &         &       &     &  imf2  & 4 pt. solid red pentagon  \\
               &                  &           &   100   &       &     &  standard  & 4 pt. solid black hexagon  \\
               &                  &           &         &       &     &  no fallback  & 4 pt. solid blue hexagon  \\
               &                  &           &         &       &     &  imf2  & 4 pt. solid red hexagon  \\
               &                  &           &   135   &       &     &  standard  & 4 pt. solid black heptagon  \\
               &                  &           &         &       &     &  no fallback  & 4 pt. solid blue heptagon  \\
               &                  &           &         &       &     &  imf2  & 4 pt. solid red heptagon  \\
               &                  &           &   50    &       &     &  Kroupa13  & 4 pt. blue cross  \\
200000         &    10, 11, 12    &    69.0   &   10    &   10  & 100 &  standard   & 3 pt. solid black triangle  \\
               &                  &           &         &       &     &  no fallback  & 3 pt. solid blue triangle  \\
               &                  &           &         &       &     &  imf3  & 3 pt. solid red triangle  \\
               &                  &           &   20    &       &     &  standard  & 3 pt. solid black square  \\
               &                  &           &         &       &     &  no fallback  & 3 pt. solid blue square  \\
               &                  &           &         &       &     &  imf3  & 3 pt. solid red square  \\
               &                  &           &   10    &       &     &  Kroupa13  & 3 pt. blue cross  \\
100000         &    10, 11, 12    &    69.0   &   10    &   10  & 100 &  standard   & 2 pt. solid black triangle  \\
               &                  &           &         &       &     &  no fallback  & 2 pt. solid blue triangle  \\
               &                  &           &         &       &     &  imf3  & 2 pt. solid red triangle  \\
               &                  &           &   20    &       &     &  standard  & 2 pt. solid black square  \\
               &                  &           &         &       &     &  no fallback  & 2 pt. solid blue square  \\
               &                  &           &         &       &     &  imf3  & 2 pt. solid red square  \\
50000          &    10, 11, 12    &   37.96   &   20    &   10  & 100 &  standard   & 1 pt. solid black triangle  \\
               &                  &           &         &       &     &  no fallback  & 1 pt. solid blue triangle  \\
               &                  &           &         &       &     &  imf2  & 1 pt. solid red triangle  \\
               &                  &           &   25    &       &     &  standard  & 1 pt. solid black square  \\
               &                  &           &         &       &     &  no fallback  & 1 pt.  solid blue square  \\
               &                  &           &         &       &     &  imf2  & 1 pt. solid red square  \\
               &                  &           &         &       &     &  kick2  & 1 pt. solid green square  \\
               &                  &           &         &       &     &  no fallback +  imf2  & 1 pt. solid magenta square  \\
               &                  &           &         &       &     &  kick2 + imf2 & 1 pt. solid cyan square  \\
               &                  &           &   30    &       &     &  standard  & 1 pt. solid black pentagon  \\
               &                  &           &         &       &     &  no fallback  & 1 pt. solid blue pentagon  \\
               &                  &           &         &       &     &  imf2  & 1 pt. solid red triangle  \\
               &                  &           &   25    &       &     &  Kroupa13  & 1 pt. blue cross  \\
\hline
\end{tabular}
\label{table:initial-MC}
\end{table*}

\subsection{$N$-body models:  NBODY6} \label{nbody6}

We use the NBODY6 direct $N$-body code \citep{aarseth03} to evolve a series of model clusters to an age 
of 12 Gyr.  We list here only those model assumptions directly relevant to the tidal field, and refer the 
reader to \citet{webb13} for more specific details regarding the input parameters 
(e.g. metallicity, binary orbital distributions, etc.).  Every $N$-body model begins with 96000 stars and 4000 
binaries (i.e. an initial binary fraction of 4\%)\footnote{We use the $N$-body models to quantify the effects 
of Galactic tides only.  Hence, for our purposes, the results are approximately insensitive to the initial 
binary fraction, which is the same in all models.}, a total 
initial mass of $6 \times 10^4$ M$_{\odot}$ 
and a half-mass radius of 6 pc, but follows a different orbit through the Galaxy.  In 
particular, only the initial velocity changes between models, giving rise to 
different orbits within the Galactic potential.  

To study the effect of a non-static tidal field, we simulate model clusters with orbital 
eccentricities of 0.5 and 0.9, each with a perigalactic distance of 6 kpc.  To help quantify any 
differences between static and non-static tidal fields, we also simulate clusters 
with circular orbits at the perigalacticon and apogalacticon of each eccentric model.  This produces 
models with circular orbits at 6 kpc, 18 kpc, and 104 kpc.  Note that model names are based on orbital 
eccentricity (e.g. e09), and the distance at apogalacticon (e.g. r104).

The initial conditions for our $N$-body models have been summarized in 
Table~\ref{table:initial-Nbody}.  In the column labeled ''Model'', we distinguish 
between simulations corresponding to different orbits by providing the eccentricity (e), 
and either the orbital semi-major axis (R$_{\rm c}$) if the orbit is circular or the 
perigalacticon distance (R$_{\rm p}$) if the orbit is eccentric.

\begin{table*}
\caption{Initial conditions for all $N$-body (NBODY6) models.}
\begin{tabular}{|c|c|c|c|c|c|}
\hline
Total Number of Stars  &   Time (in Gyrs)   &  r$_{\rm t}$ (in pc)  &  $f_{und}$   &   Model &  Symbol   \\
\hline
100000 & 12 & 40 & 6.7 & e = 0 + $R_c$ = 6 & solid black line \\
100000 & 12 & 40 & 6.7 & e = 0.5 + $R_p$ = 6 & solid blue line \\
100000 & 12 & 90 & 15.0 & e = 0 + $R_c$ = 18 & solid red line \\
100000 & 12 & 40 & 6.7 & e = 0.9 + $R_p$ = 6 & solid green line \\
100000 & 12 &  120 & 20.0 & e = 0 + $R_c$ = 104 & solid magenta line \\
\hline
\end{tabular}
\label{table:initial-Nbody}
\end{table*}



The clusters follow orbits within a Galactic potential modeled by a $1.5 \times 10^{10} M_{\odot}$ 
point-mass bulge, a $5 \times 10^{10} M_{\odot}$ \citet{miyamoto75} disk (with $a=4.5\,$kpc and $b=0.5\,$kpc), 
and a logarithmic halo potential \citep{xue08}, as described in \citet{aarseth03} and \citet{praagman10}. The 
combined mass profiles of all three components give 
rise to a circular velocity of 220 km/s at a galactocentric distance of $8.5\,$kpc.  All clusters are 
made to orbit within the plane of the disk to eliminate the effects of tidal heating due to a non-spherically 
symmetric field and disc shocking.

\section{Results} \label{results}

In this section, we present the results of our simulations for globular cluster 
evolution.  We begin by comparing the results of all simulations to the observed 
distributions of concentration, MF slope, binary fraction, and integrated V-band 
cluster magnitude.  This is done to assess the overall agreement between our models 
and the observations.  Next, we quantify the degree to which each of the dynamical 
mechanisms listed in the previous section could have contributed to the observed 
c-$\alpha$ relation.  

In order for the comparisons to be meaningful, it is 
crucially important that the simulated cluster properties are calculated analogously 
to the observed values.  In other words, it is necessary to ``observe'' the simulated
clusters in the same way as was done for the observations.  \textit{For the 
remainder of this paper, we calculate the concentration parameter using the 
cluster half-light radius instead of the tidal radius}, since the latter can be
ambiguous, particularly in the models, resulting in an ambiguous definition of
the concentration parameter \citep[e.g.][]{demarchi10}.  We call this the 
\textit{half-light concentration parameter}, denoted by 
$c_{\rm h} = \log(r_{\rm h}/r_{\rm c})$.  Both the core and half-light radii are 
calculated from the 2-D surface brightness profiles of the models, and the 
core radius is defined as the distance from the cluster centre at which the 
surface brightness falls to half its central value.

We further ensure that
the mass function slope and binary fraction are consistently calculated over the
range of stellar masses (0.3 - 0.8 M$_{\odot}$) and binary mass ratios (q $>$ 0.5) 
used to derive the observed values.  All f$_{\rm b}$ values refer to the \textit{core} 
binary fractions.  
Finally, in analogy with the observed MFs, we consider binaries as unresolved single
stars in the models when calculating the MF slope, with luminosities and colours
determined by the combined light of the binary components \citep{kroupa91}.

All models presented in this section were performed using the MOCCA code, with 
the exception of Section~\ref{gal-tides2} for which only models performed using the NBODY6 code 
are presented.

\subsection{Comparisons to the observed distributions} \label{all}

We show the results for all Monte Carlo models in Figures~\ref{fig:c-alpha-obs} 
and~\ref{fig:Mv-fb-obs} after 12 Gyr of cluster evolution, along with the observed 
values for comparison.  Specifically, the \textit{open red circles} show the observed values 
taken from \citet{demarchi07}, supplemented with additional global MF slopes taken from
\citet{paust10} for those clusters not included in the study of \citet{demarchi07}.  
We do not show our model results after 10 
and 11 Gyr of cluster evolution to avoid over-populating Figures~\ref{fig:Mv-fb-obs} 
and~\ref{fig:c-alpha-obs}.  However, if all snapshots at 10, 11 and 12 Gyr are 
included, which somewhat reproduces the age spread in the Milky Way GC population, 
the agreement with the observed distributions is slightly better, and our 
over-arching conclusions are unaffected.

The first plot shows in the c$_{\rm h}-\alpha$ plane the observed values of 
\citet{demarchi07} with our 
simulated values, whereas the second shows in the M$_{\rm V}$-f$_{\rm b}$ plane the 
observed values of \citet{milone12} with our simulated values.  Binary fractions apply 
only to the cluster core, and only to mass ratios q $> 0.5$.  
Integrated V-band magnitudes M$_V$ are calculated from the final 
total cluster luminosity assuming M$_{V,\odot} = 4.83$.

For the \textit{filled} symbols, the different colours correspond to different 
assumptions pertaining to the IMF and stellar remnants.  In particular, the black,
red, green, blue, cyan and magenta colours correspond to the standard IMF (imf1),
modified Kroupa IMF (imf2), zero BH kick velocity (kick2), no BH fallback (no
fallback), modified Kroupa IMF with zero BH kick (imf2+kick2) and modified Kroupa IMF
with no BH fallback (imf2+no fallback) models, respectively.

In an effort to better communicate to the reader the results presented in 
Figures~\ref{fig:c-alpha-obs} and~\ref{fig:Mv-fb-obs}, we also systematically vary 
the size and shape of each symbol.  For the filled points and crosses, 
the increasing point sizes correspond to increasing initial numbers of stars, with
$N=50000$, $N=100000$, $N=200000$, $N=300000$ and $N=1800000$.  The initial ratio
f$_{und} =$ r$_{\rm t}$/r$_{\rm h}$ also increases with increasing number of sides
(i.e. triangle, square, pentagon, hexagon, etc.) for symbols of a given colour
(see Table~\ref{table:initial-MC} for the exact f$_{und}$ values).  

The \textit{filled} points correspond to models with an initial binary fraction of 10\%.  
All \textit{open} symbols (neglecting the red open circles) correspond to models with 
larger initial binary fractions.  Specifically, the black, blue, and green open symbols 
correspond to 
initial binary fractions of 30\%, 70\% and 95\%, respectively.  All models corresponding 
to these open symbols adopt $N=300000$ stars initially and a standard Kroupa IMF (imf1).  
For these open symbols, 
the increasing point sizes correspond to increasing maximum binary orbital separations,
where we consider the values a$_{\rm max} =$ 100, 200 and 400 AU.  The black cross
corresponds to a standard
$N=300000$ model with an initial binary fraction of 10\%, but with initial binary
mass segregation imposed.  Finally, the blue crosses 
(labeled Kroupa13 in Table~\ref{table:initial-MC}) correspond to our standard model (imf1) with an initial binary fraction
of 10\%, but adopting the initial binary orbital parameter distributions of
\citet{kroupa13}.

\subsubsection{The observed c$_{\rm h}-\alpha$ relation} \label{obs-calpha}

\begin{figure}
\centering
\includegraphics[width=\columnwidth]{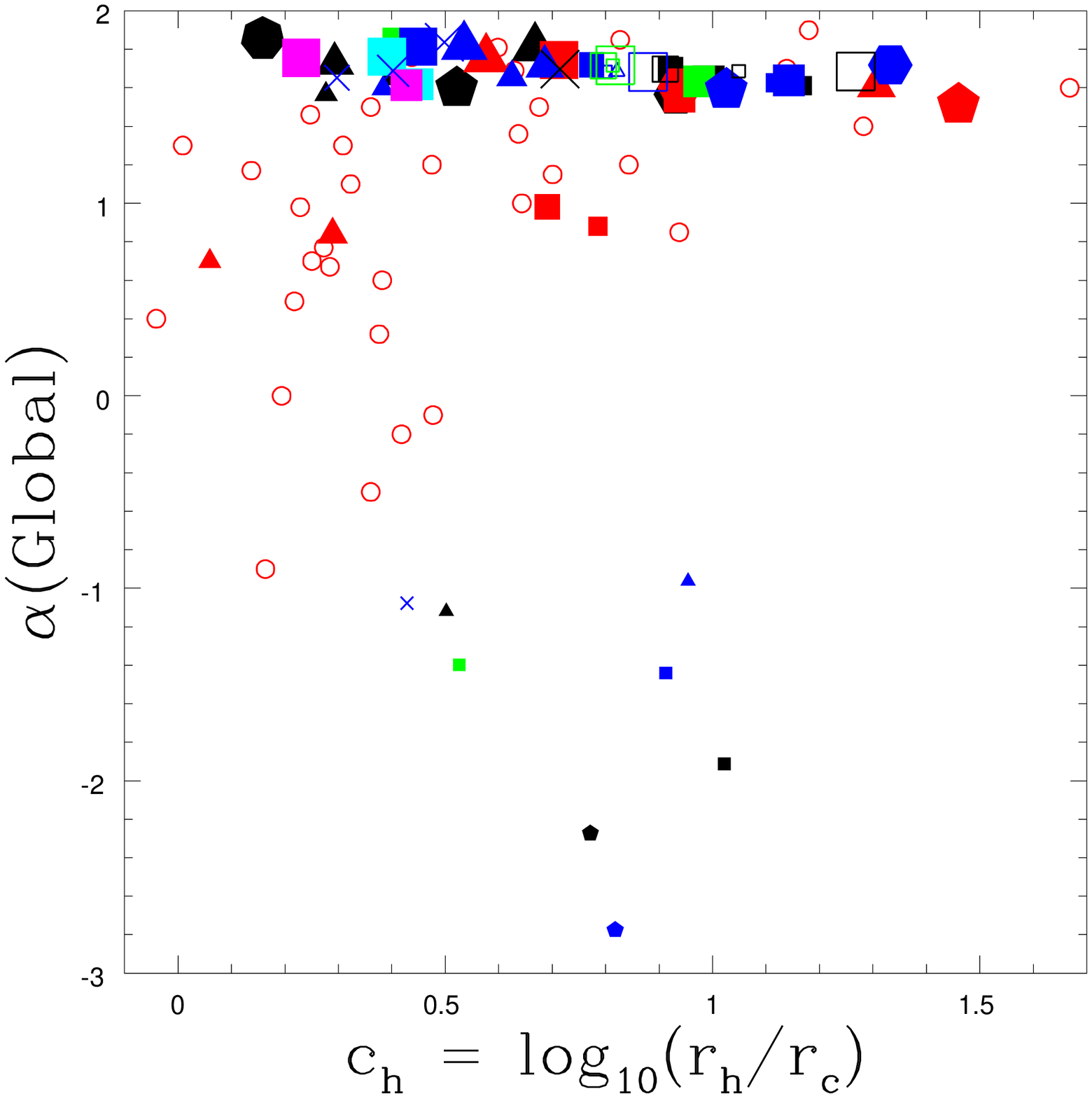}
\caption{Results for all models in the c$_{\rm h}$-$\alpha$-plane at 12 Gyr, where 
$\alpha$ is the power-law index of the MF over the mass range 0.3 - 0.8 M$_{\odot}$.  
The half-light concentration is defined as c$_{\rm h} = \log$(r$_{\rm h}$/r$_{\rm c}$).  
A full description of each coloured symbol is provided in the text.  The open red 
circles show the observed values 
taken from \citet{demarchi07}, supplemented with additional global MF slopes taken from
\citet{paust10}.}
\label{fig:c-alpha-obs}
\end{figure}

Figure~\ref{fig:c-alpha-obs} shows that the simulated ranges in the concentration 
parameter and $\alpha$ do not completely agree with the observations.  
Specifically, the models struggle to reproduce \textit{both} 
low-concentration low-$\alpha$ clusters and high-concentration high-$\alpha$ 
clusters.  
Hence, if we increase (decrease) the initial 
concentration in all models, we will struggle to reproduce clusters with 
sufficiently low (high) concentrations without imposing additional assumptions (e.g. 
the formation of an IMBH).  
We struggle to reproduce sufficiently low-$\alpha$ values in low-concentration
clusters in models that assume a standard Kroupa IMF (imf1), which is the case for 
most of our models.  As illustrated by the solid red triangles and squares in 
Figure~\ref{fig:c-alpha-obs}, the agreement appears 
better in models that assume a two-segment Kroupa IMF (imf2), and would be
better still assuming an even more bottom-heavy IMF than considered in this paper.  

As shown by the smallest points in Figure~\ref{fig:c-alpha-obs}, there is a large gap in 
$\alpha$ between the $N=50000$ models and the rest, with the $N=50000$ models also 
under-predicting $\alpha$.  This can be corrected by adopting a slightly younger 
age for these clusters.  However, most of the $N=50000$ 
models are approaching disruption at 12 Gyr, since they have lost a considerable fraction 
of their initial mass.  Hence, the rate of escape of stars across the tidal boundary 
is high, as is the internal rate of two-body relaxation.  
At 11 Gyr, some of the $N=50000$ clusters have low-$\alpha$ values and low 
concentrations, in rough agreement with the observations.  However, at 12 Gyr, all of these 
clusters have fully dissolved.  Hence, this is a short-lived phase of the cluster evolution, 
lasting less than 1 Gyr.  
Consequently, it seems unlikely that \textit{all} of 
the low-$\alpha$ low-concentration clusters observed in \citet{demarchi07} are in the 
process of fully dissolving.

\subsubsection{The observed M$_{\rm V}$-f$_{\rm b}$ relation} \label{obs-fbMv}

\begin{figure}
\centering
\includegraphics[width=\columnwidth]{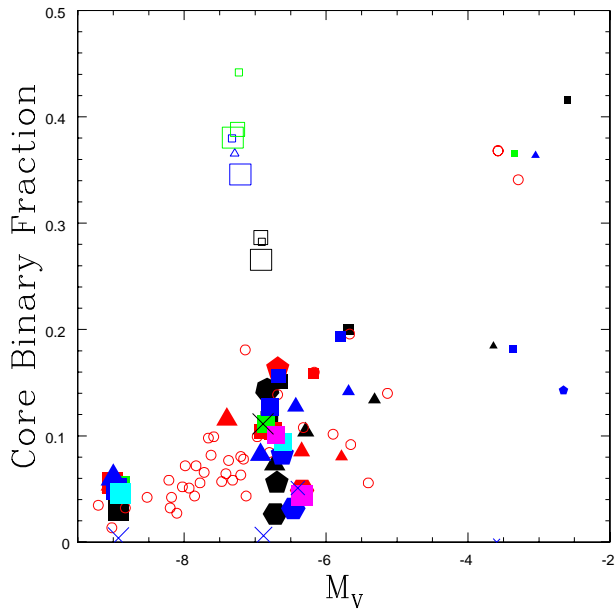}
\caption{Results for all models in the M$_V$-f$_b$-plane at 12 Gyr.  The 
colour-coding as well as the properties (size and number of sides) of the data 
points are the same as in Figure~\ref{fig:c-alpha-obs}.  The open red circles 
show the observed values taken from \citet{milone12}.}
\label{fig:Mv-fb-obs}
\end{figure}

As is clear from Figure~\ref{fig:Mv-fb-obs}, the simulated ranges in both M$_{\rm V}$ 
and f$_{\rm b}$ are in excellent agreement with the observations for the majority of 
our models.  This suggests that our assumed range in the distribution of initial 
cluster masses is 
in reasonable agreement with that of the initial \textit{cluster} mass function for
the sample of \citet{milone12}.  Additionally, \textit{for the assumed binary 
orbital parameter distributions adopted in this paper, the data are well-reproduced 
assuming an universal initial binary fraction of $\approx$ 10\% in all clusters, 
independent of the cluster mass}.  This is supported by the fact that models with 
initial binary fractions $f_{\rm b} =$ 30, 70 and 95\%, shown by the open squares 
in Figure~\ref{fig:Mv-fb-obs}, all over-predict the final 
binary fraction at the present cluster age.  We obtain good agreement with the 
observations independent of whether or not we assume initial binary mass 
segregation (shown by the black cross in Figure~\ref{fig:Mv-fb-obs}).  

Models that assume the initial binary orbital parameters distributions of 
\citet{kroupa13} and an initial binary fraction of 10\%, shown by the blue 
crosses in Figure~\ref{fig:Mv-fb-obs}, yield final 
binary fractions that are lower than, but still comparable to, our 
other models and the 
observed values.  This is because, for the initial cluster densities adopted 
in our models, a very large fraction of the initial binaries are soft, and 
are hence rapidly destroyed.  This can be corrected by truncating the initial 
period distribution closer to the hard-soft boundary, so that a larger fraction 
of the initial binaries are hard, and hence more resilient to dynamical 
disruption.  In this case, the distributions in \citet{kroupa13} would also 
yield good agreement to the observed binary fractions.  Alternatively, 
the same result could be achieved by increasing the initial binary fraction, so 
that it is higher in denser clusters with hard-soft boundaries corresponding to 
very short orbital periods.  In general, we are unable to rule out the 
possibility that other combinations of the initial binary fraction and orbital 
parameter distributions could also be consistent with the observations, such as 
the binary universality hypothesis described in \citet{kroupa11a}.  This will be 
the focus of a forthcoming paper.  

\subsection{The initial conditions} \label{initial}

\begin{figure}
\centering
\includegraphics[width=\columnwidth]{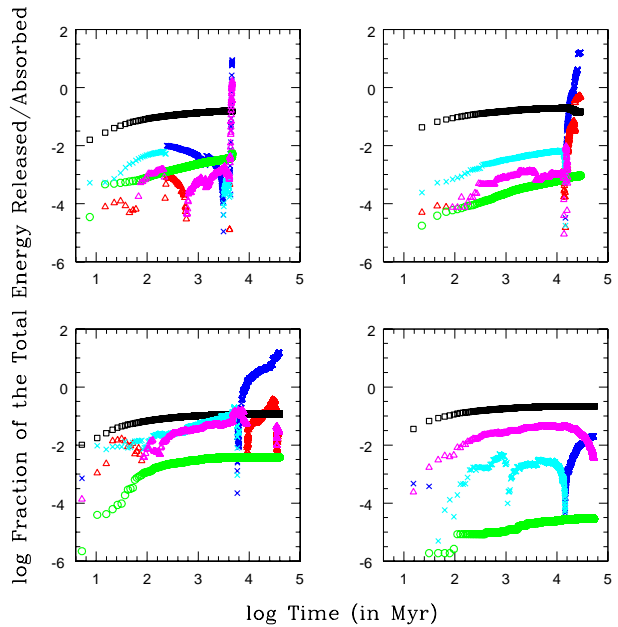}
\caption{Heating and cooling as a function of time (in Myr) for a few selected 
models.  On the y-axis, the fraction of the total initial binding energy 
of the cluster that is absorbed or released by each mechanism is plotted.  
The dark blue crosses, red triangles and 
black squares correspond to heating due to single-binary encounters, 
binary-binary encounters and stellar evolution, respectively.  The cyan crosses, 
magenta triangles and green circles correspond to cooling due to single-binary encounters, 
binary-binary encounters and direct stellar collisions, respectively.  
Models shown in the top 
insets begin with N $=$ 1800000, f$_{\rm b} =$ 0.1, a$_{\rm max} =$ 100 AU and 
either f$_{\rm und} =$ 60 (right) or f$_{\rm und} =$ 125 (left).  The model 
shown in the bottom left inset begins with N $=$ 300000, f$_{\rm b} =$ 0.1, 
a$_{\rm max} =$ 100 AU and f$_{\rm und} =$ 100.  Finally, the model shown in 
the bottom right insets begins with N $=$ 300000, f$_{\rm b} =$ 0.95,
a$_{\rm max} =$ 400 AU and f$_{\rm und} =$ 35.}
\label{fig:heat-cool}
\end{figure}

In this section, we describe how our results depend on each of the key assumptions 
that go into defining the initial conditions for our simulated clusters.  We 
begin by quantifying in Figure~\ref{fig:heat-cool} the efficiency of the various 
energy equipartition-driven 
mechanisms for cluster heating and cooling over the course of the cluster 
lifetime.  We will refer to Figure~\ref{fig:heat-cool} throughout the 
subsequent sub-sections, in which we consider these mechanisms 
individually in more detail, and how they affect the final half-light 
concentration parameter c$_{\rm h}$ and MF slope $\alpha$ .


Figure~\ref{fig:heat-cool} shows as a function of time the degree of cluster 
heating due to single-binary encounters (blue crosses), binary-binary encounters (red triangles) 
and stellar evolution-driven mass loss (black circles), as well as the degree of cooling 
due to single-binary encounters (cyan crosses), binary-binary encounters (magenta triangles) and 
direct stellar collisions (green circles).  The MOCCA code divides model clusters 
into radial bins.  Hence, the energy released/absorbed due to heating/coolng is 
calculated in each bin at each time-step, summed over the entire cluster, and divided 
by the total initial binding energy of the cluster (excluding the internal binding energy of 
binaries).  Time-steps are taken at $\sim 7.5$ Myr intervals.  The 
energy released/absorbed due to stellar evolution-driven mass loss is calculated 
from the change in potential energy, whereas the energy 
released/absorbed due to dynamical interactions is calculated directly 
from the Fewbody output.  

As an example, consider the cyan crosses in Figure~\ref{fig:heat-cool}, 
which show the degree of cooling due to single-binary encounters.  In the lower right panel, 
cooling due to single-binary interactions initially amount to $\sim 10^{-6}$ of the 
total initial binding energy of the cluster.  This increases with time to a level of 
$\sim 10^{-2}$ of the initial cluster binding energy at $\sim 1$ Gyr.  The fractional energy 
absorbed due to single-binary encounters remains roughly constant (apart from a temporary dip at 
$\sim 1$ Gyr) at $\sim 10^{-3}-10^{-2}$ for the next $\sim 10$ Gyr.  During this time, the 
long-period binary fraction is reduced due to disruption.  It finally reduces to zero at 
$\sim 11-12$ Gyr when all soft binaries have been disrupted.  After this, single-binary 
encounters primarily serve to further harden close binaries, acting as a heat source by 
imparting energy to single stars (shown by the dark blue crosses in 
Figure~\ref{fig:heat-cool}).

In all models, stellar evolution (black squares) is the 
dominant heating mechanism early on in the cluster lifetime, when massive stars are 
still present in significant numbers.  After several Gyr of cluster evolution, 
single-binary (dark blue crosses) and, to a lesser extent, binary-binary (red triangles) encounters 
take over as the dominant heating source.  Early on in the cluster lifetime, however, 
both single-binary (cyan crosses) and binary-binary (magenta triangles) encounters act as an 
important source of cooling, since soft binaries are still present in significant numbers.  
Finally, direct stellar collisions (green circles) are never 
the dominant source of cooling, and tend to absorb energy at a much lower rate than 
the other heating/cooling mechanisms.

\subsubsection{Binaries} \label{soft2}

The disruption of soft binaries is only effective as a cooling mechanism for the first $\sim$ 
few Gyr of cluster evolution.  However, during this time, the cooling due to soft binary 
disruption is outweighed by heating due to stellar evolution-induced mass loss, since massive 
stars are still present in significant numbers.  The heating caused by the mass loss from these 
massive stars contributes to an expansion of the core, and the energy sink provided by the 
disruption of soft binaries only serves to slow the mass loss-driven expansion of the core.  This is 
shown in Figure~\ref{fig:heat-cool}.  Cooling due to the disruption of soft binaries is shown 
by the cyan crosses and magenta triangles, whereas heating caused by stellar evolution-driven mass 
loss is shown by the black squares.

Clusters that begin with high initial 
binary fractions do not necessarily end up with high concentrations, independent 
of our assumption for the maximum orbital separation, and hence the fraction of soft 
binaries.  The open squares in Figures~\ref{fig:c-alpha-obs} and~\ref{fig:Mv-fb-obs} 
correspond to model clusters with high initial binary fractions.  As is clear from 
Figure~\ref{fig:c-alpha-obs}, all of these models end up with final core binary fractions 
that are much higher than observed.  At the same time, Figure~\ref{fig:c-alpha-obs} 
shows that these same clusters end up with final concentrations that are approximately 
independent of the initial binary fraction.

It seems unlikely that the disruption of soft binaries contributed significantly
to the observed distribution of concentration parameters.  In order for the disruption 
of soft binaries to have a significant effect 
on the core radius, and hence concentration, our results suggest that the period 
distribution would need 
to be heavily peaked just beyond the hard-soft boundary.  In this case, most binaries 
would be soft, with orbital energies slightly larger than that corresponding to the 
hard-soft boundary.  This would maximize the effectiveness of soft binary disruption 
as an energy sink, and would help to create massive clusters with high 
concentrations.  However, we are unaware of a theoretical reason for why the period 
distribution should be heavily peaked near the hard-soft boundary, and this would 
conflict heavily with empirically-derived period distributions 
\citep{kroupa95,kroupa11b}.  

Binary burning 
becomes an effective heat source much later in the cluster lifetime, typically 
only after $\sim 10$ Gyr of evolution (with the exception of the model shown in 
the upper left panel in Figure~\ref{fig:heat-cool}, which evolved to a high 
central density on a shorter time-scale than the other models).  
This is because, for most of the lifetime of a cluster, the concentration is 
increasing, and sufficiently high central densities are required in order for 
interactions between (primarily) single stars and binaries to occur at a fast enough rate 
for binary burning to become effective.  In our models, binary burning mainly 
serves to slow the contraction of the core, as opposed to completely reversing the 
collapse and driving a re-expansion of the core.  The effectiveness of binary 
burning is shown by the 
blue crosses and red triangles in Figure~\ref{fig:heat-cool}, which correspond to heating 
due to single-binary and binary-binary encounters, respectively.

\subsubsection{Galactic tides} \label{gal-tides2}



\begin{figure}
\centering
\includegraphics[width=\columnwidth]{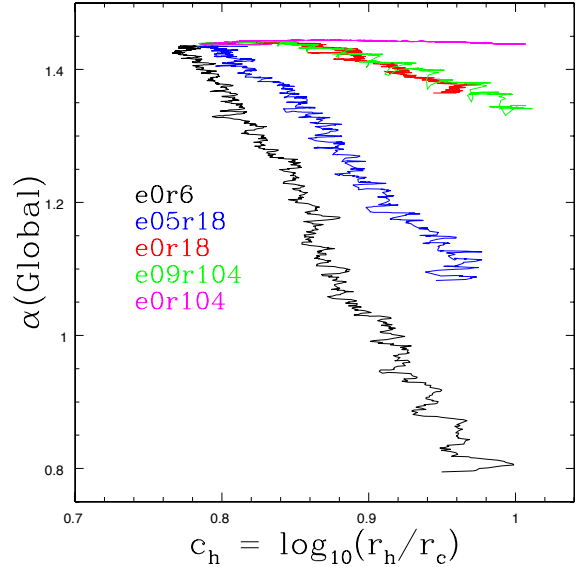}
\caption{Evolution in the c$_{\rm h}$-$\alpha$ plane over 12 Gyr for each $N$-body model.  All 
clusters begin with $\alpha \approx +1.5$ and evolve toward slightly higher concentrations 
and lower MF slopes.  Each model is colour-coded, such that model names are based on the distance at 
apogalacticon r$_{\rm a}$ (e.g. r104 for r$_{\rm a} =$ 104 kpc), and the orbital
eccentricity (e.g. e09 for e $=$ 0.9).  Note that the observed trend \citep{demarchi07} is 
opposite to that evident here.}
\label{fig:nbody2}
\end{figure}

In this section, we rely almost exclusively on the results of our $N$-body models to 
quantify the effects of Galactic tides on the observed c$_{\rm h}$-$\alpha$ relation.  
These are presented in Figure~\ref{fig:nbody2}, which shows the evolution in the 
c$_{\rm h}$-$\alpha$-plane for models with different orbits through the Galaxy but 
identical initial conditions (see Section~\ref{nbody6} for the specific initial 
conditions).  The black, blue, red, green, and magenta lines, in that order, correspond 
to orbits with decreasing average tidal fields.  The main conclusion to be drawn from 
Figure~\ref{fig:nbody2} is that \textit{Galactic tides typically have only a small effect on the 
evolution of the ratio c$_{\rm h} =$ r$_{\rm h}$/r$_{\rm c}$, but a large effect on 
the evolution of the global stellar MF.}  Below, we explain the origin of this important 
result.  We will refer back to Figure~\ref{fig:nbody2} throughout this section to 
help illustrate our results.

Galactic tides affect the evolution of the stellar MF in the following way.  For a given 
cluster mass, clusters that experience the strongest tidal fields undergo the most rapid mass 
loss due to the fact that the tidal radius decreases with decreasing Galactocentric distance.  
Consequently, clusters exposed to stronger tidal fields undergo 
the most rapid flattening of their MFs (i.e. the greatest rate of decrease in $\alpha$), since 
low-mass stars are preferentially accelerated to wider orbits within the cluster potential.  
This mass loss translates into a reduction in the time-scale for two-body relaxation.  
The shorter relaxation time exacerbates the trend, so that more 
mass is lost from the cluster at an ever-increasing rate, shortening the relaxation time even 
further.  Thus, on average, we expect clusters that experience the strongest 
tidal fields to undergo the most rapid flattening of their stellar MFs.

Galactic tides affect the evolution of the cluster concentration in the 
following way.  There is an overall trend for clusters that experience the strongest tidal fields to 
have the smallest radii \citep{webb13}.\footnote{In this section, 
we always refer to the 3-D radii, instead of the observational values calculated from the 
surface brightness profiles.}  This trend sets in 
within roughly the first Gyr of evolution, and becomes exacerbated as the clusters 
continue to evolve.  This is because clusters with small Galactocentric distances (and hence 
small tidal radii) have the least room to expand before filling their tidal radii.  Once 
these clusters are tidally-filling, their core and half-mass radii begin to contract 
due to two-body relaxation \citep[e.g.][]{spitzer87,heggie03}.

Despite the fact that all radii 
change considerably over the cluster evolution, the net effect of 
these changes is that the evolution in the concentration parameter is small when 
calculated using r$_{\rm h}$ instead of r$_{\rm t}$, changing by a factor only 
slightly greater than unity.  \textit{Although r$_{\rm c}$ and r$_{\rm h}$ 
themselves change significantly, the ratio 
r$_{\rm h}$/r$_{\rm c}$ remains roughly the same for all clusters}.  This is clearly 
illustrated in Figure~\ref{fig:nbody2} 
for all simulated clusters, almost independent of their orbit through the Galaxy.  That is, 
for model clusters with the same initial size and concentration the evolution in c$_{\rm h}$ 
appears to be orbit-independent in Figure~\ref{fig:nbody2}, with c$_{\rm h}$ increasing 
from 0.8 to almost 1 after a Hubble time.  Any differences are within the observational 
uncertainties.  The 
only exception to this is initially tidally under-filling clusters on relatively circular 
orbits at large Galactocentric distances.  In these clusters, the core radius 
decreases due to two-body relaxation at 
a noticeably faster rate than does the half-mass radius.  This contributes to an 
increase in the concentration over time while $\alpha$ remains more or less constant, 
albeit the effect remains small in our models, even for the most distant orbits (see below).  

To better explore the effect that initially tidally under-filling clusters may have on
the results of 
Figure~\ref{fig:nbody2}, we make use of lower mass versions of our $N$-body models.  These lower
mass models contain 48000 single stars and 2000 binaries initially, and are described in detail
in \citet{webb13}.  Due to their smaller masses, these models are less computationally expensive,
which allows us to explore a range of initial half-mass radii.  For clusters with initial
half-mass radii of 2 pc and 4 pc, the central concentration increases while the cluster
expands to fill its tidal radius (see Figure 7 in \citet{webb13}), and $\alpha$ remains 
roughly constant.  This results in tidally 
under-filling clusters evolving towards the high-c$_{\rm h}$, high-$\alpha$ region of
Figure~\ref{fig:c-alpha-obs}.  Once the tidal radius is filled, $\alpha$ decreases at a rate 
similar to that in the models shown in Figure~\ref{fig:nbody2}, while maintaining a near 
constant half-mass concentration c$_{\rm h}$.  However, in \textit{very} tidally under-filling 
clusters at large 
Galactocentric distances, 12 Gyr may not be enough time to enter this phase
of evolution.  Thus, although we are unable to reproduce the highest concentrations observed
by \citet{demarchi07}, initially very tidally under-filling clusters tend to produce the highest
final concentrations.  


\textit{These results suggest that Galactic tides should contribute 
to, and likely even dominate, the dispersion in the observed c$_{\rm h}-\alpha$ relation.}  
Most of this dispersion should appear at the 
low-concentration (and typically low-mass) end of the distribution, since these include both 
initially massive, heavily stripped clusters at small Galactocentric distances and 
initially low-mass clusters at large Galactocentric distances that have lost only a 
small fraction of their mass.  This can help to account for some of the dispersion 
in $\alpha$ at the low-c$_{\rm h}$ end not reproduced in our models, as 
seen in Figure~\ref{fig:c-alpha-obs}.

More quantitatively, the smallest Galactocentric 
distance in the sample of \citet{demarchi07} is $\gtrsim 2.1$ kpc and $\alpha$ 
does not evolve significantly at Galactocentric distances beyond $\approx 100$ kpc.  Hence, 
at the low-concentration end of the distribution, we estimate from 
Figure~\ref{fig:nbody2} that tides should contribute to a dispersion in $\alpha$ of 
$\delta{\alpha} \approx \pm (0.6-0.7)$ \textit{for a given initial cluster mass}.  This 
value of $\delta{\alpha}$ is a minimum, since the clusters plotted in 
Figure~\ref{fig:nbody2} all began with the same initial total mass.  Hence, $\delta{\alpha}$ 
should be larger for a range of initial total cluster masses, as expected in the 
proto-Milky Way \citep[e.g.][]{marks12,kroupa13}.  For comparison, the presently 
observed range is $\delta{\alpha}_{\rm obs} \approx 3.0$ (from -1.0 up to 2.0).  Given the 
range of initial cluster masses and Galactocentric distances that should apply to the 
sample of \citet{demarchi07}, we conclude that Galactic tides dominate the dispersion 
in the observed distribution of present-day MF slopes. 

We note that tides contribute significantly 
to altering the \textit{global} mass function, however the effect is much less 
pronounced for the central MF evaluated near the core (see Webb et al., in preparation).  
This is because the \textit{central} MF is primarily altered via two-body relaxation-driven 
mass segregation, and is relatively insensitive to the escape of stars across the tidal 
boundary.  
This explains why the results presented 
in \citet{leigh12}, which focused on the annulus immediately outside the core (between 
one and two core radii from the cluster centre), are consistent with the general picture 
that the cluster-to-cluster differences observed in the PDMFs of GCs arise from 
an universal IMF modified primarily by two-body relaxation-driven dynamical evolution.  
This general picture is also consistent with the results presented in this paper.

\subsubsection{The initial stellar and remnant mass functions} \label{remnants2}

\begin{figure}
\begin{center}
\includegraphics[width=\columnwidth]{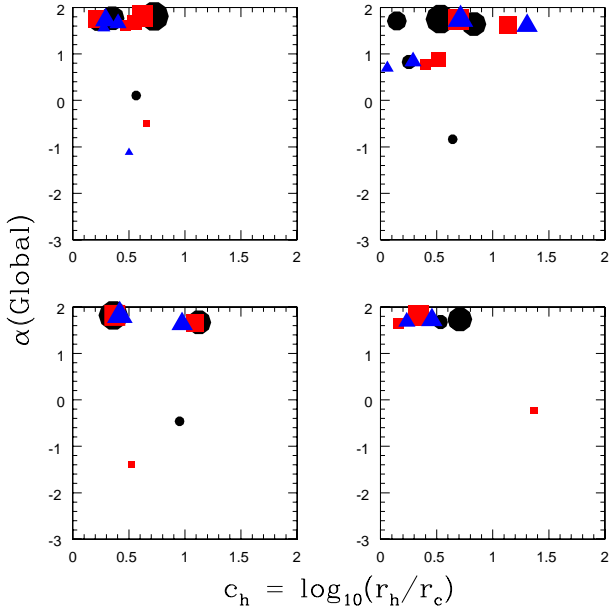}
\end{center}
\caption[$c_{\rm h}-\alpha$ plot for different IMFs and initial-final mass relations 
for BHs]{Results in the $c_{\rm h}-\alpha$-plane for different IMFs and initial-final 
mass relations for BHs.  We show the results for a normal Kroupa IMF (top left inset), 
as well as an IMF with a single break mass at 0.5 M$_{\odot}$ and low-mass and 
high-mass slopes of +1.3 and +2.3, respectively (top right inset).  We also 
try adjusting the amount of fallback onto BHs upon their formation.  The no 
fallback case is shown in the bottom left inset.  Finally, the bottom right 
inset shows our results assuming both a non-standard two-segment IMF and mass 
fallback onto BHs.  
\label{fig:rem}}
\end{figure}

In Figure~\ref{fig:rem}, we show how our results change upon varying both 
the initial stellar mass function and the initial-final mass relation for 
BHs.  The black circles, red squares, and blue triangles correspond to 
snapshots at 10, 11 and 12 Gyr, respectively.  In order of increasing 
size, the size of the points correspond to clusters with 50000, 100000, 
200000, 300000 and 1800000 stars initially.

As illustrated by the black (10 Gyr), red (11 Gyr) and blue (12 Gyr) squares in 
Figure~\ref{fig:rem}, the slope 
of the MF tends to decrease over time.  Clusters with the lowest masses (i.e. $N=50000$; 
shown by the smallest point-size in Figure~\ref{fig:rem}) end up with the smallest MF 
slopes, in qualitative agreement with the observations \citep[e.g.][]{paust10,leigh12}.  
More quantitatively, Figure~\ref{fig:c-alpha-obs} shows that models that assume a standard 
Kroupa IMF struggle to reproduce 
clusters with sufficiently low-$\alpha$ and low-concentration, as was found by 
\citet{zonoozi11} for the GC Palomar 14.  The agreement is slightly better in models that 
assume a non-standard IMF, as shown by the solid red triangles and squares in 
Figure~\ref{fig:c-alpha-obs}.  
However, this is only because the non-standard IMF has more massive 
stars, which lose the most mass due to stellar evolution, and this contributes 
to an expansion of the core.  Thus, the final concentration is smaller for 
a non-standard IMF due to the additional mass loss from massive stars very 
early on in the cluster evolution.  

As mentioned, we confirm that the half-light concentration tends 
to increase over time, however this need not always be the case.  For example, 
stellar evolution causes an expansion of the core very early on in the cluster 
lifetime, and the concentration can change from increasing to 
decreasing during the final stages of cluster dissolution.  More importantly, 
\textit{there is a stochasticity in the simulations such that nearly identical initial 
conditions can produce significantly different final evolutionary states for 
our model clusters.}  

As illustrated in Figure~\ref{fig:rem}, most of the stochasticity observed in 
our Monte Carlo simulations seems to be tied to the dynamical 
evolution of the remnant sub-population within the cluster.  For example, in some 
of our models, an IMBH forms through a new pathway not yet discussed in the 
literature (for more details, see Giersz, Leigh \& Hypki 2013, in preparation).  In 
this scenario, an IMBH forms from a single originally stellar-mass black hole that grows 
in mass due to dynamical interactions that induce mergers between the growing BH and 
(typically) other stellar remnants, combined with binary evolution-driven mass-transfer 
events.  To initiate this process, one BH must be left in the system after all 
BH-forming supernovae have ceased, or a single BH must be formed via NS-NS or 
NS-WD mergers.  The presence of additional BHs tends to prevent the formation of a 
single very massive BH due to competitive merging and accretion, followed by their 
dynamical ejection from the cluster.  
The process is facilitated by the fact that the growing BH is rarely 
isolated, since the time-scale for it to capture another object and form a binary 
is very short.  The presence of such a binary companion is crucial, since it reduces 
the time-scale for dynamical interactions and hence mergers.  In our standard models, 
the mass growth of the BH is typically slow 
and requires a few Gyr of cluster evolution before an IMBH forms.  
The final core radius is typically higher in these models 
than in those for which no IMBH forms.  

Therefore, our results show that remnants can also contribute 
to the dispersion in the observed c-$\alpha$ relation.  Based on the results 
shown in Figure~\ref{fig:rem}, we estimate that remnants can contribute to a 
dispersion in $c_{\rm h}$ of up to $\delta{c_{\rm h}} \approx 1.0$, and possibly 
more in a few rare cases.

\section{Discussion} \label{discussion}

In this section, we argue that stellar dynamics alone is not enough to 
reproduce the observed c$_{\rm h}-\alpha$ relation.  While the effects of two-body 
relaxation combined with Galactic tides \textit{can} reproduce the observed dispersion 
in $\alpha$ \citep{leigh12}, we do not find 
a strong increase in the concentration parameter with increasing cluster mass, and hence 
increasing $\alpha$ \citep[e.g.][]{demarchi10,leigh12}, in our models, regardless 
of the initial (universal) conditions and assumptions adopted.  This implies that some 
other non-dynamical mechanism is also required to reproduce the observed distribution 
of concentration parameters.  That is, the initial distribution of cluster concentrations 
cannot be universal when clusters begin evolving in relative isolation due solely to energy 
equipartition-driven dynamical evolution.  Some other mechanism that is independent of the 
internal dynamical evolution of clusters is also required.  We go on to explore the possibility 
that the origin 
of the observed c$_{\rm h}-\alpha$ relation is tied to the gas-embedded phase 
of cluster evolution.

\subsection{Energy equipartition-driven dynamical evolution alone cannot reproduce 
the observed c$_{\rm h}-\alpha$ relation}

Based on our results, two-body relaxation-driven dynamical evolution alone cannot 
reproduce the observed c$_{\rm h}-\alpha$ 
relation.  This is because our simulated clusters struggle to 
\textit{simultaneously} produce both high-concentration high-$\alpha$ and 
low-concentration low-$\alpha$ clusters, provided we adopt the same 
empirically-motivated universal IMF for every cluster \citep{kroupa11a,kroupa13}.  
The problem is that, unlike the IMF, we do not have a reasonable empirically- or 
theoretically-motivated guess at what the 
initial concentration should be, as a function of any cluster parameter.  

Shifting the initial mass function slope and/or distribution of 
concentrations in either direction will only worsen the agreement at the 
opposite end.  For example, if we begin with clusters that are initially 
more concentrated, we will struggle even more to produce sufficiently 
low concentration clusters with flat MFs, and vice versa.  However, in 
principle, it is possible to match the observed c$_{\rm h}-\alpha$ relation at the 
high-c$_{\rm h}$ high-$\alpha$ end by assuming a higher initial concentration.  
Similarly, a lower initial 
concentration combined with a lower value for $\alpha$ would improve the 
agreement at the low-c$_{\rm h}$ low-$\alpha$ end.

For these reasons, it is difficult to constrain the exact spread in the initial 
distribution of concentration parameters required to explain the observed 
distribution.  Instead, we re-iterate that the observed spread in $\alpha$ 
\textit{can} be reproduced from an universal IMF combined with energy 
equipartition-driven dynamical evolution \citep{leigh12}.\footnote{Although we have 
shown consistency with the universality hypothesis for the IMF, our results do 
not serve as a proof that the universality hypothesis is correct.  Although variations 
in the IMF with the initial cluster mass (or other cluster properties) have not been 
ruled out in this study, any such dependences explored in future studies 
should be firmly rooted in star formation theory given that the observations are 
also in general consistent with the universality hypothesis for the IMF 
\citep[e.g.][]{kroupa13}}.  This quantitatively 
reproduces the general trend that lower mass clusters tend to have shallower 
PDMFs \citep{demarchi07,demarchi10,paust10}.  Meanwhile, the 
observed spread in c$_{\rm h}$ \textit{cannot} be reproduced from an universal 
distribution of initial concentrations.  Thus, some other mechanism that is 
independent of the internal dynamical evolution of clusters must also be 
operating in order to reproduce the (weak) trend that higher mass clusters 
(with, on average, steeper MFs) tend to have higher present-day concentrations 
\citep{harris96}.  In the 
subsequent sections, we explore different possibilites that could contribute 
to this observational trend, with a focus on the gas-embedded phase of cluster 
evolution.

\subsubsection{Caveat:  stellar collisions and tidal interactions} \label{tides}

\citet{milgrom78} argued that tidal dissipation during close fly-bys between
single stars can reduce the stars' kinetic energies, and contribute
to a contraction of the core on a shorter time-scale than is achieved
by two-body relaxation alone.  Similarly, direct collisions between
single stars should also dissipate kinetic energy \citep{lightman77}.
The single-single collision rate increases with increasing cluster
mass \citep{leonard89}, so that the most massive clusters should experience
the most efficient tidal dissipation due to close fly-bys and direct
collisions between single stars.  Thus, tidal interactions and collisions
may both contribute to the observed $c-\alpha$ relation, since they
both contribute to a correlation between cluster mass and concentration.

A detailed treatment of the effects of collisions and tidal interactions is 
beyond the scope of this paper, however our results suggest that these 
effects can likely be ignored in future studies.  As illustrated in 
Figure~\ref{fig:heat-cool}, the energy sink provided by collisions is never the 
dominant heating/cooling mechanism affecting the concentration.  In fact, the 
effect is typically negligible, and remains approximately constant over the 
cluster lifetime for the central densities reached in our models.  
It follows that the energy sink offered by tidal interactions is 
also unlikely to ever be the dominant 
heating/cooling mechanism.  This is because the rate of tidal capture events 
should be comparable to the rate of direct collisions, since the efficiency of 
tidal capture decreases rapidly with increasing distance of closest approach.  
Also, tidal interaction events that do not result in binary formation 
typically remove significantly less energy than tidal captures, since the 
energy dissipated due to tides 
quickly becomes negligible with increasing distance of closest approach.

\subsection{The gas-embedded phase of cluster formation} \label{embedded}

In the previous section, we argued that it is unlikely that the observed 
c$_{\rm h}-\alpha$ relation has a purely energy equipartition-driven origin.  It 
follows from this, and the assumption that the observed (central) MF 
distribution alone \textit{can} be 
accounted for purely by two-body relaxation-driven dynamical evolution 
\citep[e.g.][]{paust10,leigh12}, that the origin of the observed distribution of 
cluster concentrations must be tied to some other physical process(es).  For 
example, hierarchical merging of clusters early on in their lifetimes could 
perhaps contribute.  Alternatively, external perturbations from the Galaxy may 
also play a role, such as disc shocking due to passages through the plane of 
the Galaxy or interactions with nearby giant molecular clouds or other star 
clusters.  In this section we focus on the gas-embedded phase of cluster 
formation, and discuss some of the various mechanisms that could have operated 
when gas was still present in significant quantities.


\subsubsection{Cluster expansion due to rapid gas expulsion} \label{expulsion}

\citet{marks08} argue that rapid gas expulsion combined with primordial 
mass segregation can produce the observed low-concentration clusters 
with flat MFs.  Clusters expand in response to the sudden removal of their 
gas.  Although the tidal radius also increases due to the loss of mass, the 
central cluster regions expand more significantly, causing the concentration 
to decrease.  At 
the same time, preferentially low-mass stars in the outskirts become 
unbound, and escape from the cluster.  This simultaneously lowers 
the concentration and decreases the mass function slope, improving the 
agreement at the low-c$_{\rm h}$ low-$\alpha$ end of the observed relation.
The results presented in this paper suggest that an even more bottom-heavy IMF in 
the mass range 
0.3 - 0.8 M$_{\odot}$ than assumed in our standard model (imf1), combined with a lower
initial half-light concentration, could reproduce the observed c$_{\rm h}-\alpha$
relation at the low-c$_{\rm h}$ end.  Importantly, this would approximately
reproduce the post-gas expulsion MF and cluster concentration described
in \citet{marks08} and \citet{marks12}.  Disc shocking and external perturbations from 
other massive bodies on nearby orbits within the Galaxy could also contribute 
to this general trend.  This is because these mechanisms should typically desposit 
additional energy 
within clusters, causing them to expand and accelerating the rate of escape of 
preferentially low-mass stars across the tidal boundary \citep[e.g.][]{vesperini97}.

\citet{marks08} also caution that unresolved binaries can contribute to making the 
MF index $\alpha$ appear smaller than it actually is.  This is because each 
binary causes two single stars to disappear from the mass function, and an additional 
star with a derived mass higher than either binary component to be included.  Hence, on 
average, binaries artificially deplete the low-mass end of the MF while simultaneously 
over-populating the high-mass end.  Unresolved binaries should have at most a small 
effect on our results, however.  This 
is because the simulated mass functions are derived by treating binaries as unresolved 
objects, in analogy with the observed mass functions.  Additionally, in all but 
the lowest mass clusters considered here, the number of binaries is so few that 
they do not significantly affect the derived power-law index of the MF 
\citep[e.g.][]{milone12}.

\subsubsection{Cluster contraction due to prolonged gas retention} \label{retention}

There now exists evidence that the most massive MW GCs underwent multiple 
episodes of star formation \citep[e.g.][]{piotto07}.  This trend does not appear in 
much lower mass open clusters, and even the lowest mass globulars 
\citep[e.g.][]{gratton12}.  The currently favoured scenario to explain these 
multiple populations suggests that either star formation was on-going 
for $\approx$ 10$^8$ years \citep[e.g.][]{conroy11}, or massive clusters 
re-accreted gas from which new stars were formed \citep[e.g.][]{pflamm09}.  This 
additional star formation indirectly implies that gas was present in these 
clusters for a prolonged period of time relative to their lower mass 
counterparts.  

Several mechanisms could contribute to increasing the concentration during 
the gas-embedded phase.  For example, the occurrence of star formation suggests 
relatively high gas accretion rates for the stars.  Independent of whether or not 
the gas actually 
remains bound to the accretor, this could act to reduce the accretor velocities 
due to conservation of momentum.  For this to be the case, all that is required 
is that the ``accreted'' gas be accelerated by the accretor such that the two 
are co-moving relative to the background medium \citep{hoyle39,bondi44,bondi52}.  
In general, accretion from the interstellar medium should both increase 
the central stellar density and accelerate the 
mass segregation process \citep{leigh13}, since the accretion rate typically 
increases with increasing accretor mass.  

Active star formation also 
suggests high gas densities.  Consequently, gas dynamical friction could 
be very efficient, since the gas dynamical friction force scales 
linearly with the gas density.  The net effect of gas
dynamical friction is to transfer kinetic energy from the stars to the surrounding
gas, causing the central stellar density to increase but the central gas
density to decrease.  We note that 
the gas dynamical friction force also depends critically on whether or not 
the motion of the perturber relative to the gas is subsonic or supersonic.  
Given that stars are actively forming during much of the gas-embedded phase, this 
suggests that the gas must have been relatively cold.  It follows that the sound-speed 
should have been small compared to the stellar velocity dispersion, and hence that the 
motion was predominantly supersonic.  The 
most efficient gas drag occurs when the relative velocity is slightly less than
the sound speed \citep{lee11} (i.e. when the stellar and gas velocity dispersions
are roughly equal), however previous work has also shown it to be effective in
the supersonic regime \citep{dokuchaev64,ruderman71,rephaeli80,ostriker99}.

The increase in stellar density due to the presence of gas could result in positive 
feedback, and help to prolong the 
gas-embedded phase.  This is because a higher stellar density translates into a 
deeper potential well, and hence a larger escape velocity.  This could in turn imply 
a larger gas retention fraction and/or a longer gas expulsion time 
\citep[e.g.][]{heggie09}.  

If more massive clusters tend to have higher gas densities primordially, retain their 
gas for longer or re-accrete more gas from the surrounding intra-cluster medium 
than do low-mass clusters, this could contribute to 
a correlation between cluster mass and concentration.  This is because, in this case, 
the efficiencies of both accretion from the ISM (onto stars) and gas dynamical friction 
should increase with increasing cluster mass.  

We conclude that the gas-embedded phase 
of cluster evolution could be crucially important for the origin of the observed 
c$_{\rm h}-\alpha$ relation.  This is not only due to the fact that the results 
presented in this paper suggest that two-body relaxation-driven dynamical evolution 
alone cannot explain the observed distribution of concentrations, but 
also because the gas damping mechanisms we have considered 
should all contribute to the observed trend of increasing concentration 
with increasing cluster mass.  More work will be needed to better isolate the 
origin of the observed c$_{\rm h}-\alpha$ relation and, in particular, the degree to which 
it was present at the end of the gas-embedded phase.  Future studies with this goal should 
aim to constrain the gas-embedded phase of cluster evolution (e.g. central gas 
density, gas retention time, etc.) as a function of the total cluster mass.

\section{Summary} \label{summary}

In this paper, we consider the origin of the correlation between cluster 
concentration and present-day mass function slope observed in the Milky Way globular 
cluster population.  To this end, we generate a suite of Monte Carlo 
and $N$-body models using the MOCCA and NBODY6 codes, respectively, for star cluster 
evolution.  We compare the results to the observed correlation between cluster 
concentration and mass function slope, as well as to the observed anti-correlation between 
cluster mass and binary fraction.  These relations can either be reproduced from universal 
initial conditions combined with some dynamical mechanism(s) that alter(s) the 
distributions over time, or they must arise very early on in the cluster lifetime, such as 
during the gas-embedded phase of cluster formation.  We explore a number of dynamical 
mechanisms that could contribute to, or even 
reproduce, the observed trends.  Our key conclusions are:

\begin{list}{*}{}
\item Galactic tides combined with energy equipartition-driven dynamical evolution can 
account for most, if not all, of the observed dispersion in the present-day MF slope $\alpha$, 
but not all of the dispersion in the concentration parameter c$_{\rm h}$.
\item Dynamical effects induced by stellar remnants should also contribute non-negligibly to 
the dispersion observed in c$_{\rm h}$.
\item Some other mechanism that is independent of the internal dynamical evolution of clusters 
must also operate in order to reproduce the observed dispersion in c$_{\rm h}$.
\item Whatever the mechanism(s) responsible, it should operate by preferentially increasing 
c$_{\rm h}$ in more massive clusters in order to reproduce the observed trend that 
clusters with steep MFs (and typically large total cluster masses) tend to have the 
highest concentrations.
\end{list}

Thus, we conclude 
that energy equipartition-driven dynamical evolution alone could not have reproduced the 
observed relation between concentration and mass function slope.  Consequently, we suggest 
that this 
trend could be connected to the gas-embedded phase of cluster evolution.  Specifically, we 
argue that cluster contraction due to prolonged gas retention could account for the most
concentrated clusters with the steepest mass functions, and cluster expansion due
to rapid gas expulsion could account for the least concentrated clusters with the
flattest mass functions.

\section*{Acknowledgments}

We would like to thank Michael Marks and Hagai Perets for useful discussions, as well as 
an anonymous referee whose suggestions helped to improve this manuscript.  This work 
was partly supported by the Polish Ministry of Science and Higher 
Education through the grant N N203 38036 and by the National Science Centre
through the grants DEC-2012/07/B/ST9/04412 and DEC-2011/01/N/ST9/06000.


\bsp

\label{lastpage}

\end{document}